\documentclass[a4paper,sigconf,screen,authorversion]{acmart}

\usepackage{amsmath}
\usepackage{subfig}
\usepackage{tikz}

\renewcommand\footnotetextcopyrightpermission[1]{} % removes footnote with conference information in first column
\setcopyright{none}
\copyrightyear{2022}
\newcommand\copyrighttext{%
  \footnotesize\sffamily Copyright \copyright 2022. This manuscript version is made available under the CC-BY-NC-ND 4.0 license.}
\newcommand\copyrightnotice{%
\begin{tikzpicture}[remember picture,overlay]
\node[anchor=south,yshift=50pt] at (current page.south) {{\parbox{\dimexpr\textwidth-\fboxsep-\fboxrule\relax}{\copyrighttext}}};
\end{tikzpicture}%
}

\begin{document}

\title[nOS-V: Co-Executing HPC Applications Using System-Wide Task Scheduling]{nOS-V: Co-Executing HPC Applications\\Using System-Wide Task Scheduling}

\author{David Álvarez}
\email{david.alvarez@bsc.es}
\affiliation{%
  \institution{Barcelona Supercomputing Center}
  \country{Spain}
}

\author{Kevin Sala}
\email{kevin.sala@bsc.es}
\affiliation{%
  \institution{Barcelona Supercomputing Center}
  \country{Spain}
}

\author{Vicenç Beltran}
\email{vbeltran@bsc.es}
\affiliation{%
  \institution{Barcelona Supercomputing Center}
  \country{Spain}
}

\begin{abstract}
Future Exascale systems will feature massive parallelism,
many-core processors and heterogeneous architectures.
In this scenario, it is increasingly difficult for HPC applications
to fully and efficiently utilize the resources in system nodes.
Moreover, the increased parallelism exacerbates the effects of existing
inefficiencies in current applications.
Research has shown that co-scheduling applications to share system nodes
instead of executing each application exclusively can increase resource
utilization and efficiency.
Nevertheless, the current oversubscription and co-location techniques to share
nodes have several drawbacks which limit their applicability and make them
very application-dependent.

This paper presents co-execution through system-wide scheduling.
Co-execution is a novel fine-grained technique to execute multiple HPC applications
simultaneously on the same node, outperforming current state-of-the-art
approaches. We implement this technique in nOS-V, a lightweight tasking library
that supports co-execution through system-wide task scheduling. Moreover,
nOS-V can be easily integrated with existing programming models, requiring no changes
to user applications.
We showcase how co-execution with nOS-V significantly reduces schedule makespan for several
applications on single node and distributed environments, outperforming prior node-sharing techniques.
\end{abstract}

% \lhead{Copyright}

\maketitle

\fancyhead[L]{\sffamily\footnotesize{Submitted to the Journal of Parallel and Distributed Computing}}
\fancyhead[R]{\sffamily\footnotesize{David Álvarez, Kevin Sala and Vicenç Beltran}}
\fancyfoot[C]{\vspace{1em}\sffamily\thepage}
\copyrightnotice

% \thispagestyle{empty}

% HERE GO THE PAPER SECTIONS

\section{Introduction}

Traditionally, HPC clusters have been partitioned into subsets of nodes that execute applications exclusively, without being shared
for the duration of the application's execution.
This differs from other workloads like cloud computing, which strive to share single cluster nodes with as many users as possible.

% The exclusive execution model is efficient as long as HPC applications can exploit all their allocated nodes.
The exclusive execution model is efficient as long as HPC applications can fully exploit all their allocated nodes.
However, typical HPC workloads can display various issues that prevent them from
efficiently using a whole node. % a whole node => each node during all execution time.
For example, most applications have serial initialization, finalization or even communication phases,
where there is limited parallelism.
Fork-join or distributed applications can display load imbalance due to an ineffective distribution
of available work. % at the node or cluster level, respectively => this sentence is mixing too many things, load imbalance can be at the cluster or node level, but it doesn't really depende on fork-join or tasks model.
Some applications even lack parallelism when scaling to nodes with hundreds of cores.
Additionally, with the rise of heterogeneous systems in HPC, some applications cannot overlap accelerated
and non-accelerated computations, leaving resources idle.

Moreover, the end of Dennard scaling~\cite{1050511} and the rise of many-core systems and heterogeneous machines exacerbate the aforementioned issues.
Exascale systems not only have to worry about scalability and resilience problems but also the decaying node-level efficiency
hindering total cluster-level efficiency.

Co-scheduling applications on the same node has been proposed as a way to mitigate the problems mentioned above.
These techniques modify cluster schedulers (e.g. SLURM~\cite{SLURM}) to identify applications to co-schedule and execute simultaneously on the same nodes.
The main goal is to increase available parallelism by running more than one application, filling the parallelism gaps in one application with work from another.
However, resources must be shared fairly and efficiently amongst applications.

HPC applications make use of runtime systems that divide their workloads into discrete work units (tasks), which can be executed in parallel.
Runtimes assume they are running exclusively in each node, and they schedule tasks over the system's resources.
These presumptions of exclusivity over the node's resources and sparse preemptions by the OS allow runtimes to apply aggressive optimizations such as busy waiting or custom synchronization techniques.
In practice, this causes that system noise preempting runtime threads can significantly influence application performance.

Therefore, while in non-HPC workloads we could simply oversubscribe the applications on the same node and let the underlying Operating System (OS) perform time-sharing,
this approach does not work well in the case of HPC applications.
Due to their reliance on exclusive access, HPC applications can have strong interferences when executed simultaneously~\cite{DBLP:conf/nsdi/BasetWT12, tangram, impactover, oversubs2, oversubs3}.
This can result in co-scheduled workloads displaying unpredictable behavior and significant slowdowns.
Two well-known problems contribute to this behavior.
First, a large number of created threads can cause a scalability collapse scenario~\cite{collapse1, collapse2}.
Second, the OS preempting threads that rely on user-space spinlocks can cause the widely-known Lock Holder Preemption and Lock Waiter Preemption problems~\cite{lhp1, lhp2, lhp3}.

For HPC workloads, static co-location~\cite{coloc-hpc,bubble-flux,slurm-drom,coloc-case} has been proposed to reduce application interference.
With static co-location, system resources are statically partitioned, and each application runs confined in a specific partition.
This approach simulates each application running exclusively on a subset of the node.
However, this fails to adapt to the dynamic nature of HPC applications, which display different amounts of parallelism at different points in time.

We propose a new technique called \textit{application co-execution},
which can improve overall node efficiency by leveraging the instant parallelism of all co-scheduled applications.
With co-execution, we maintain a global view of the available tasks from all applications running in the system.
This is more similar to the OS's view, but work can be scheduled in a suitable granularity for HPC applications, preventing oversubscription problems.

We implement this approach through the novel nOS-V tasking library.
nOS-V is a portable library that leverages widely available inter-process communication (IPC) facilities and requires no kernel modifications.
When using nOS-V for co-execution, users can execute several HPC applications simultaneously in the same node.
However, a single runtime (nOS-V) manages system resources, instead of one per application.
Additionally, nOS-V supports global scheduling policies based on process priority or data locality.

Specifically, our contributions in this work are:
\begin{enumerate}
	\item We present the technique of application co-execution for HPC workloads
  	\item We describe nOS-V, a novel tasking library that enables application co-execution and
 	\item We evaluate in detail the performance of nOS-V integrated with the OmpSs-2 task-based programming model
 	and show how it outperforms other node-level resource sharing techniques such as oversubscription and co-location
\end{enumerate}

% Certainly space for a background would be nice
% Main section
\section{Background}

Several solutions to tackle and improve node-level efficiency and increase
job throughput in HPC clusters have been studied for a while.

\begin{figure*}
    \centering
    \includegraphics[width=1\textwidth]{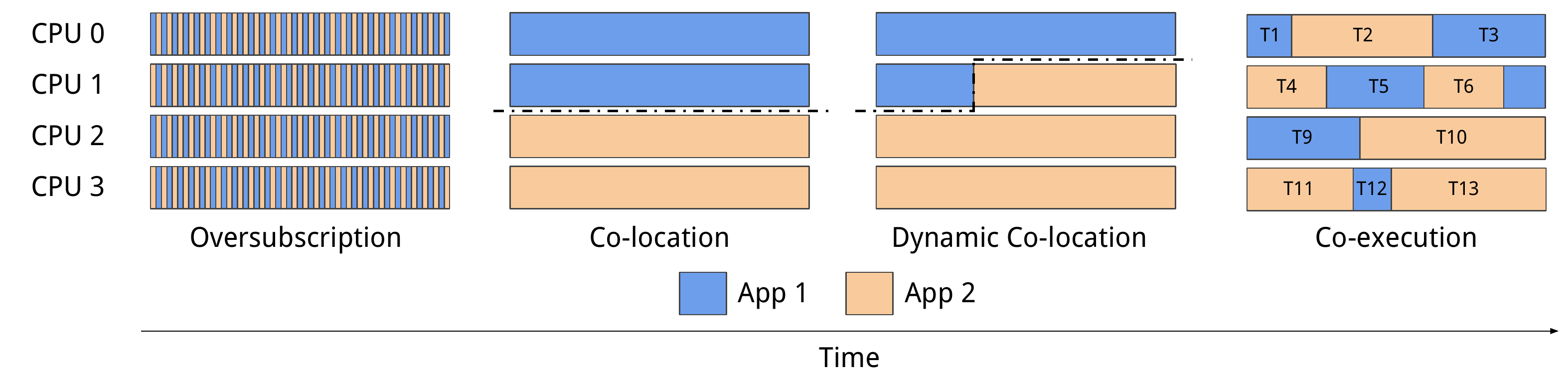}
    \caption{Diagram of two applications co-scheduled using oversubscription, co-location, dynamic co-location and co-execution}
    \label{fig:fig-all}
\end{figure*}

The simplest solution is known as oversubscription, which is shown in Figure~\ref{fig:fig-all}.
Although oversubscription is extensively used for Cloud workloads~\cite{DBLP:conf/nsdi/BasetWT12}, the use of
busy-waiting policies~\cite{DBLP:conf/iwomp/0001HLE16} and user-space locking~\cite{9095224} can cause
significant slowdowns on co-scheduled HPC applications~\cite{tangram, impactover, oversubs2, oversubs3}.
In essence, the slowdowns are caused by the OS pre-empting HPC runtimes while holding locks, busy waiting or performing
critical communication.

Modern approaches are usually based on application co-location, shown in the second place of Figure~\ref{fig:fig-all}.
Co-location executes independent applications in the same node by splitting system resources at the
level of core or socket and assigning each application to an exclusive set of resources.
Co-location can work well if the applications to run concurrently in the same node are
chosen carefully and the partitions are correctly sized. However, finding the ideal partition
for a set of applications may not be trivial or even feasible, as many applications feature dynamic parallelism
in distinct phases.

Dynamic co-location techniques extend static co-location with dynamic resource sharing between partitions~\cite{drom,dlb}.
An example of dynamic co-location is DLB~\cite{dlb}, which leverages malleability in
the application's programming models and acts as a broker that oversees how applications
lend and reclaim cores to each other, as shown third in Figure~\ref{fig:fig-all}.
Essentially, dynamic co-location improves over traditional co-location by changing
the static partition between processes at some points during runtime.

However, dynamic co-location does not have a global view of the available work on the node
and only oversees how applications voluntarily yield resources.
Thus, dynamic co-location cannot take informed node-wide scheduling decisions.
Moreover, like static co-location, there is a limit on the number of applications that can
be executed simultaneously, as the initial resource partitions need to have at least one core.

All approaches to improve node-level efficiency are translated into cluster-level
improvements by fitting batch schedulers like SLURM~\cite{SLURM} with job co-scheduling
capabilities that consider how node-level resources are shared
\cite{hpc-batch,slurm-drom, coloc-hpc, smite, heracles,bubble-flux}.
The co-scheduling problem is related to node-level efficiency: better alternatives
to in-node resource sharing can be combined with existing scheduling policies
to deliver better cluster-level efficiency.

In this work, we focus on HPC applications, which rely on programming
models such as OpenMP~\cite{openmp51} and OmpSs-2~\cite{bsc2020ompss2}.
Those programming models are, in turn, based on lower-level threading
or tasking libraries, often used in the implementation of high performance
programming models~\cite{threadreview,bolt}.
These tasking libraries provide lightweight parallel work units referred to as user-level threads, tasks or tasklets.

Basing programming model's runtimes upon these tasking libraries provides
several benefits over the traditional POSIX thread interface~\cite{POSIX}.
User-level threads are lighter weight and can be complemented with customizable
user-space schedulers that implement custom policies.
Several available state-of-the-art tasking libraries implement ULTs,
including Argobots~\cite{argobots}, Qthreads~\cite{qthreads} and MassiveThreads~\cite{massivethreads}.

Usually, a tasking library assumes exclusive access to the entire node and executes tasks in every core, regardless of
any other processes running in the system.
Execution is generally done by creating a pool of pthreads pinned to specific cores and scheduling tasks as
if they were being directly mapped to physical cores.
% custom scheduling policies, data locality is just one example.
This ultimately allows implementing scheduling policies which maximize data locality.

Co-execution is the main contribution of this paper, where we execute several applications simultaneously
by \textit{sharing the same tasking library instance}, as opposed to performing oversubscription or co-location.
This way, a single scheduler is shared by multiple applications,
while maintaining the property of having only one Pthread running per system core at any time. % bound ... => running per core at any time.

The novel nOS-V tasking library implements co-execution by sharing a single scheduler located
in a shared memory region containing tasks from multiple processes and, if enabled, from different users.
The fourth diagram of Figure~\ref{fig:fig-all} illustrates this concept by showing two applications with
nOS-V enabled tasking runtimes, which use nOS-V to share system resources.
Resources are not statically partitioned into sets, and applications' tasks can still
be executed on any core.
Moreover, the application runtimes may be different, as long as they are integrated with nOS-V.
By sharing the tasks to be scheduled and applying a node-wide scheduling policy, we can
guarantee that there is no oversubscription because there is no scenario where two tasks
are concurrently scheduled to the same core.

Note that this is similar to the operating system's scheduling but done
on a task granularity instead of a thread granularity, allowing task-based applications to not suffer from the
drawbacks of either oversubscription or co-location.
Additionally, as this shared scheduler lives in user-space, it can be fitted with customized
scheduling policies.

\section{The \lowercase{n}OS-V Tasking Library}
\label{sec:nanosv}

nOS-V is a new tasking library that provides a simple interface to create and manage tasks, similar to other user-level thread libraries.
The key feature of nOS-V is that it enables the co-execution of different applications on a node through a novel
inter-process tasking mechanism.
Unlike other tasking libraries that implement context-switching from user-space, our library leverages
Pthreads to context-switch threads that belong to different applications.

Like other user-level thread libraries, nOS-V provides an abstraction layer
based on tasks that hides the complexities of many-core management,
thread bindings and cooperative task scheduling. However, in nOS-V, there is only
one instance of the library managing all the cores in the node.
This abstraction layer can replace the low-level Pthreads~\cite{POSIX} API in applications and task-based runtimes.
In this way, tasks from different applications can be transparently co-executed.

nOS-V manages pools of Pthreads located in shared memory for each process being co-executed.
These threads are pinned to specific cores using \texttt{sched\_setaffinity}, and allow nOS-V to execute tasks on specific cores by running them on these threads.
When a task finishes, its Pthread is re-used to execute other tasks.
However, if a task blocks, its associated Pthread is blocked with it.
This allows nOS-V tasks support any feature supported by a regular Pthread, such as Thread Local Storage (TLS), unlike tasking libraries based on user-level context switching.
The trade-off of using Phtreads that remain attached to tasks is a higher context-switch cost only when a task blocks or yields.
However, this extra overhead is usually negligible~\cite{cpu-elasticity}, and it ultimately enables inter-process tasking.

\subsection{Architecture}

\begin{figure}
    \centering
    \includegraphics[width=1\columnwidth]{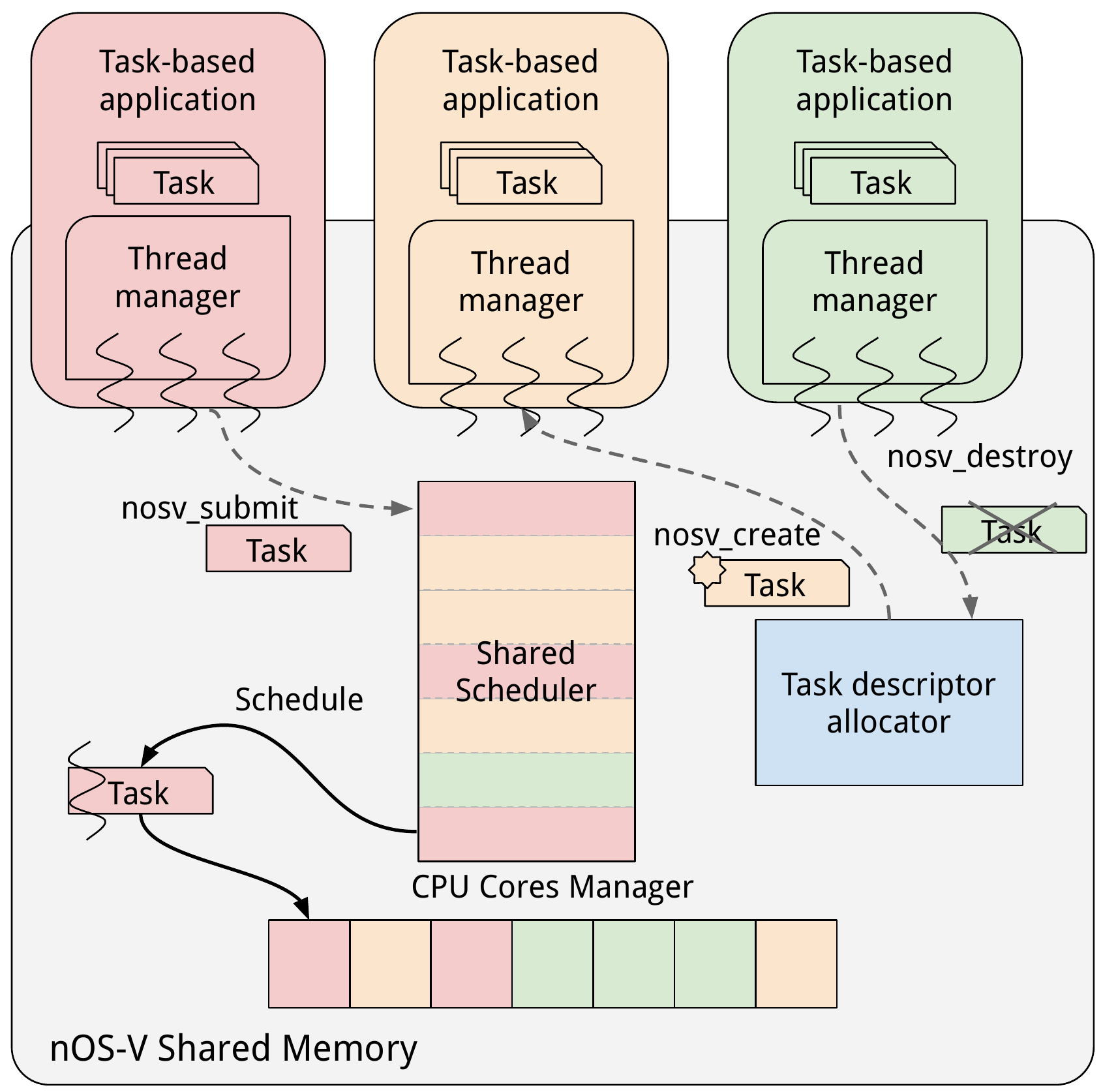}
    \caption{Architecture of nOS-V running with three attached applications}
    \label{fig:arch}
\end{figure}

Figure~\ref{fig:arch} displays the architecture of nOS-V, which will be referenced throughout this section.
The main difference between nOS-V and other tasking libraries is that most of nOS-V's components
are allocated on a POSIX Shared Memory~\cite{POSIX} segment, which is then mapped onto all processes
in the node using the library.
Shared memory is then the primary IPC mechanism used throughout our tasking library.
It has a very low overhead, does not require system calls to communicate between processes,
and allows re-using regular data structures without having to design ad-hoc alternatives.

% We could have chosen other alternatives such as Process-in-Process (PiP)~\cite{pip}, which would also have solved
% context switching between processes. However, PiP cannot work with programs from different users,
% and even if a workaround were found, there would be security concerns as processes would have shared their
% whole virtual address space.

As shown in Figure~\ref{fig:arch}, we allocate the following data structures in the nOS-V's shared memory segment:

\begin{itemize}
    \item A thread manager for every nOS-V enabled process running.
    \item A CPU manager that stores the current mapping of cores to threads.
    \item The task descriptors.
    \item A centralized scheduler based on a Delegation Ticket Lock~\cite{advancedsynchronizationtechniques}.
\end{itemize}

In the example of Figure~\ref{fig:arch}, three task-based
applications, each in a different process, are running simultaneously.
The nOS-V library will manage thread pools for each application, which
can submit tasks for execution.
Submitted tasks will be placed in a shared scheduler (see Section~\ref{sec:shared-scheduler}), containing work from
multiple applications.
Later on, when the nOS-V CPU manager has an idle core not running any task,
it will grab a ready task from the scheduler and execute it on an available POSIX
thread from the task's creator process on that idle core.

\subsection{Tasking Interface}
The main API of nOS-V contains four basic operations to handle tasks coming from
multiple processes: \texttt{nosv\_create}, \texttt{nosv\_submit}, \texttt{nosv\_pause} and \texttt{nosv\_destroy}.

The \texttt{nosv\_create} operation provides to the caller a task descriptor on which
all the relevant information to the task can be stored. The three main items that
this task descriptor contains are: the PID of the process on which the task was created,
a callback to run the task, and a callback to notify when the task has finished its execution.
nOS-V users can also embed a metadata pointer in this task descriptor to implement task arguments
or data dependency information.

The \texttt{nosv\_submit} operation inserts a previously created task in the shared scheduler and marks
it as ready for execution. The task will then remain in the scheduler until there is an idle
core and the current scheduling policy decides to execute this task. The task is executed by
calling the run callback in the descriptor in a thread that belongs to the same PID as the task descriptor.

The \texttt{nosv\_pause} operation blocks a currently running task.
This call can be used to yield the current core if a task needs to perform a blocking operation or wait for an event.
The operation must be called from the blocked task's
context, as in a yield operation. Once a task is paused, its associated Pthread is blocked and
remains attached to the task. The core where this task was running will be marked as idle, and it will become available
to run other ready tasks from the scheduler. Unblocking a task is done by calling the \texttt{nosv\_submit} operation again,
which will insert it back into the scheduler.

The \texttt{nosv\_destroy} operation returns a task descriptor to nOS-V and frees the associated shared memory. This
operation has to be called after the task has finished its execution.

Figure~\ref{fig:arch} showcases how these operations behave in the overall nOS-V architecture.
The first application performs a submit operation, which inserts the task into the shared scheduler.
Meanwhile, the second and third applications create and destroy tasks, thus interacting with the task descriptor allocator.

\subsection{nOS-V Life cycle}

Whenever an application linked with nOS-V is executed, the library
checks during startup for the existence of a specific POSIX shared memory segment~\cite{POSIX} on the node and initializes
the segment if it does not exist.
This shared memory is then mapped into the process' address space and will be used to communicate
with other nOS-V enabled processes.

The first process registered into this shared memory region spawns a new thread for each core in the node,
ready to execute available work.
Each spawned thread is pinned to a specific idle core and will ask the scheduler for work until a task is obtained or the thread is shut down.
If the obtained task belongs to a different process (the PID in the descriptor is different), the current thread is suspended, and a thread belonging to the ready task's process is obtained from its thread pool.
The selected thread is then pinned to the current core and woken up with the assigned task.
Note that tasks are always executed in a thread belonging to the process that created them.

When a task is paused through \texttt{nosv\_pause}, the Pthread currently executing the task is blocked in a condition variable.
Then, another Pthread (spawned or recycled from an idle thread pool) is resumed on the current core and executes other ready tasks.
When the paused task becomes ready again, it is put back into the scheduler's ready queue, so eventually, it will resume its execution.
When a worker thread gets a ready task from the scheduler, it checks, before executing it, if the task already has an attached thread.
If so, the current worker thread is blocked and added to a pool of idle threads.
Then, the worker thread attached to the task is unblocked, and the task can finally resume its execution in the current core.
Note that the woken-up thread's affinity is set before unblocking it, effectively performing a context switch between both threads.

Finally, a nOS-V application finishes when all the tasks have been executed.
When this happens, the process is unregistered from the shared memory structures, and the last process to unregister will delete the whole shared memory segment.

\subsection{Shared Scheduler}
\label{sec:shared-scheduler}
The shared scheduler implementation we have used is a centralized scheduler based on
a Delegation Ticket Lock (DTLock)~\cite{advancedsynchronizationtechniques}. We chose this design
because, unlike work-stealing, we can implement node-wide scheduling policies with a consistent
view of all the tasks in the node while maintaining state-of-the-art performance.

One consideration built into the scheduler was that whenever several nOS-V applications are
running, it is desirable to minimize the number of context switches caused by giving a task to a thread
belonging to a different process.
To this end, the scheduler prioritizes assigning tasks with the same PID as the thread requesting a task.
However, this could result in other processes being starved of computing resources, as one PID could
monopolize all cores in a system.
To solve this problem, we included the concept of time quantum into the scheduler.
When a specific core has been executing tasks from the same process for more than a certain amount of time (a quantum),
 the scheduler will choose a task from another process on the next task-switching point.
This causes a context switch but provides a more fair share of compute time amongst different processes.
The quantum is user-configurable.

Finally, we implemented three opt-in custom scheduling policies in nOS-V, to serve as an example of how custom scheduling policies can be implemented in nOS-V.
First, nOS-V has user-configurable per-application and per-task priorities. Per-application priorities are similar to
traditional process priorities in many operating systems. Per-task priorities allow a user to prioritize the execution of specific tasks
over others, and can be used to reduce execution times in programs with critical task dependency paths.
The last implemented policy allows defining a per-task affinity through the nOS-V API.
This policy allows nOS-V users to implement locality-based scheduling, by defining a specific core or NUMA Node where a task
will be executed. Users can also choose to make this affinity strict or best-effort.
We showcase the potential of the locality policy in our experimental evaluation.

\subsection{Memory Management}
A significant part of this library's complexity comes from efficiently managing the POSIX shared memory
region and allocating data structures for all processes in the same address space.

Managing a fixed-size contiguous memory region and supporting dynamic and scalable memory
allocation is challenging but not unique to nOS-V.
In fact, it is quite similar to the task that OS kernels do to manage available physical memory.

Although many state-of-the-art and user-space oriented scalable memory allocators exist~\cite{jemalloc}, some of them
rely on \texttt{mmap}-like semantics or use static per-process metadata, making them unsuitable to use by nOS-V.
We settled on a custom memory allocator that splits the shared memory region in chunks and uses a similar
approach to the well-known SLAB allocator~\cite{slab1, slab2}, together with per-cpu chunk caches, to provide scalable memory allocation.
Our implementation proved to be competitive with other memory allocators and is much better suited for nOS-V, as it allows
to free a pointer allocated by a different process because the allocator's metadata resides in the shared memory.

\subsection{Threat Model and Security}

nOS-V is built on a threat model that assumes co-executed applications are trusted, and hence are not ill-intentioned.
This model is analogous to other libraries which leverage shared memory across applications, such as DLB, Arachne and NVIDIA's CUDA MPS~\cite{mps}.
However, this trust does not imply that applications are bug-free, only that there are no malicious agents.

Resiliency is a critical factor when designing HPC software, even when applications are trusted.
Programs are expected to operate while failures are the norm rather than an exception, and it is expected that nOS-V can recover
when one or more of the co-executed applications encounter a bug, without corrupting state or crashing other unrelated co-executed programs.

This was the main reason for sharing only one memory segment over alternatives that share the full address space of processes such as Process-in-Process (PiP)~\cite{pip}.
Additionally, we intend to explore the use of hardware features that limit the impact in nOS-V of bugs in user applications.
For example, Intel Memory Protection Keys (MPK)~\cite{mpk} or Pointer Authentication on ARMv8.3-A~\cite{pac-1, pac-2}.

With nOS-V, it is technically possible to co-execute applications belonging to different users. In HPC settings, we imagine this feature to be used
between users from the same group, where users trust each other and may share compute-hour quotas.
By default, this feature is disabled, and nOS-V instances may only co-execute applications from the same user. However, it can be activated
through a configurable option, and then different users can co-execute applications using the same nOS-V instance.

\section{Adapting HPC Runtimes to Leverage nOS-V}

In the previous sections, we have defined co-execution, and we have presented
the nOS-V tasking library that allows co-executing tasks from multiple applications in one node.
The library's main objective is to be integrated into tasking runtime systems that support
parallel programming models, such as OpenMP~\cite{openmp51} and OmpSs-2~\cite{bsc2020ompss2}. However,
the simplicity of the nOS-V library API would allow using it directly from user applications if required.

\begin{figure}
    \centering
    \includegraphics[width=.6\columnwidth]{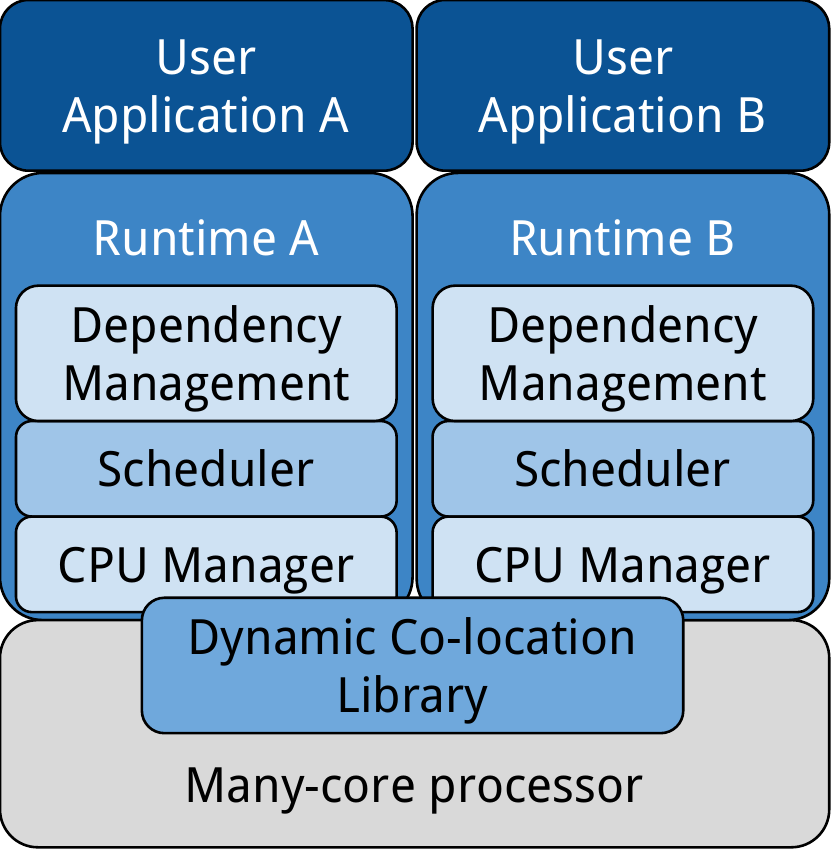}
    \caption{Diagram of two applications co-executed using Dynamic Co-location}
    \label{fig:nos-v-scheme-pre}
\end{figure}

Traditional runtimes supporting dynamic co-location work like depicted in Figure~\ref{fig:nos-v-scheme-pre}.
Two different user applications will have two different runtime instances when running in a single node.
These runtimes will have independent schedulers and \textit{CPU Managers}.
This latter is the component that tracks the available computing resources and manages their occupancy.
To support dynamic co-location, both runtimes must interact with a \textit{Dynamic Co-location Library} such as DLB,
which acts as a broker.
In this modality, runtimes must support core hot-swapping to add and remove computing resources dynamically.

\begin{figure}
    \centering
    \includegraphics[width=.6\columnwidth]{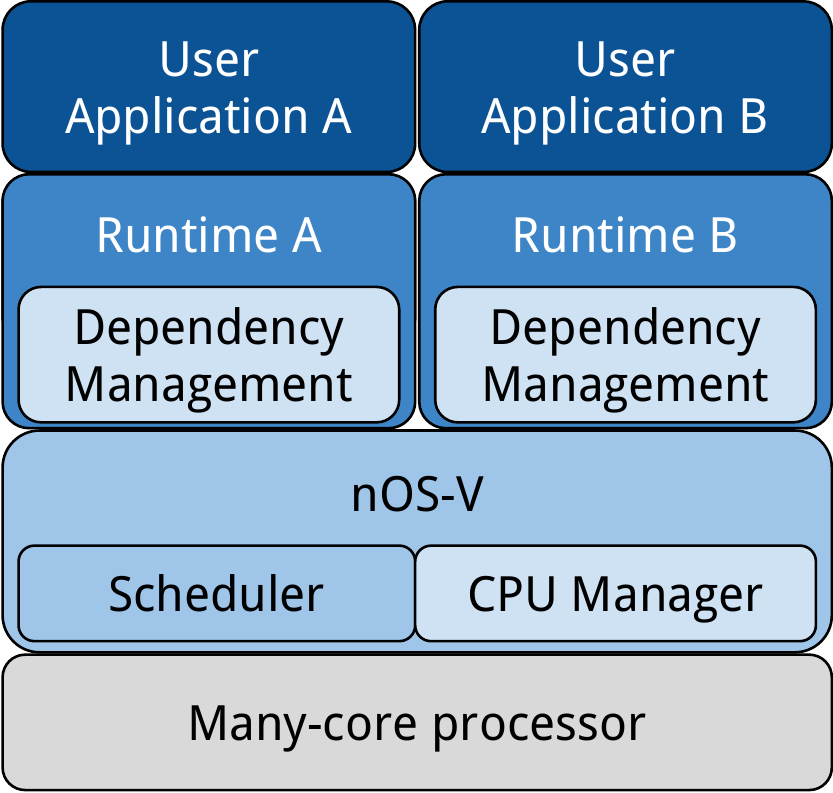}
    \caption{Diagram of two applications co-executed using nOS-V}
    \label{fig:nos-v-scheme-post}
\end{figure}

When adapting a runtime to use nOS-V, there is no need for a scheduler or a CPU manager, as the tasking library provides those components.
Instead, runtimes directly leverage nOS-V, which controls all available resources, as shown in Figure~\ref{fig:nos-v-scheme-post}.

For the experimental evaluation of this paper, we have focused on integrating the OmpSs-2 programming model, which is a data-flow task-based
programming model, similar to the OpenMP tasking model~\cite{openmp51}, but with some additional
features~\cite{reductions-ferran,WorksharingMarcos,JMWeaks,advancedsynchronizationtechniques}.
OmpSs-2 has a fully open-source implementation, and it was chosen because it is simple to extend and modify, and delivers
competitive performance with other HPC shared memory programming models.
Nanos6~\cite{bsc2019nanos6} is the reference runtime implementation for the OmpSs-2 programming model
and is linked with every OmpSs-2 program that has been compiled with either the Mercurium compiler~\cite{mercurium, mercurium2}
or a more recent LLVM-based compiler~\cite{clompss}. Nanos6 is in charge of the task creation, dependency
management, scheduling, memory allocation and CPU management of these OmpSs-2 programs.
Moreover, Nanos6 is compatible with DLB, which has a complex integration with the runtime, and will allow us to compare nOS-V to dynamic co-location.

We applied the previous modifications to the Nanos6 runtime, removing the scheduling and CPU managing components and delegating those functions to nOS-V.
The resulting implementation was used for the evaluation section of this paper.
%This process can also be applied to other HPC runtimes, allowing different implementations to leverage the same nOS-V instance.
This process can also be applied to other HPC runtimes, allowing applications using different programming models to leverage the same nOS-V instance.

\section{Experimental Evaluation}
This section shows how nOS-V minimizes the interference between applications and
improves overall resource utilization for several co-location scenarios.

For all experiments, we use the Nanos6 runtime adapted to use nOS-V.
In the first experiment, to ensure that our modified Nanos6 runtime is a suitable baseline, we will compare it against the standard
Nanos6 runtime, which has state-of-the-art performance for task-based applications~\cite{advancedsynchronizationtechniques}.
The goal is to check whether using nOS-V introduces any additional overhead when not using the co-execution capabilities,
ensuring that the changes to task allocation, scheduling or thread managing do not introduce any performance penalty.

The second experiment focuses on co-executing typical HPC applications with enough parallelism to fill all the node's cores.
We used the seven benchmarks from the first experiment and co-executed them in pairwise and three-wise combinations.

The last experiment will explore how co-execution can be leveraged for hybrid MPI+OmpSs-2 applications.
In this case, applications feature serial communication phases, and we will transparently leverage nOS-V
to provide a natural overlap of communications and computations from two distributed applications.
In addition, we will also showcase a custom scheduling policy that can be implemented in nOS-V to ensure data locality in NUMA systems.

The first two experiments were conducted on an AMD Rome cluster where each node has a single AMD EPYC 7742 64-core processor with SMT turned off and 1TiB of main memory.
The software stack is based on CentOS 8 with Linux 4.18,
Intel MKL 2021.1.1, Intel MPI 2019.2 and GCC 10.2.0.
The last experiment was conducted in an Intel Skylake cluster where each node has two sockets with Intel Xeon Platinum 8160 24-core processors with SMT off.
Nodes are interconnected with an Intel OmniPath 100Gbit network.
The software stack is based on SUSE Linux 12 with Linux 4.4, Intel MKL 2017.4, Intel MPI 2018.4 and GCC 10.1.0.
The process quantum for nOS-V is configured at 20ms for all experiments.

\subsection{nOS-V Baseline Performance}

The baseline experiment uses a benchmark set~\cite{advancedsynchronizationtechniques}
that includes a matrix multiplication, a vector dot-product, a Gauss-Seidel heat equation simulation, the HPCCG proxy application,
an N-Body simulation, a Cholesky factorization, and the Lulesh 2.0~\cite{lulesh} proxy application.

For each benchmark, we measure the performance on two points: one with an ideal task granularity which results in peak performance for
the application, and another where the task granularity is too small, and thus its performance is bound by the overhead
introduced by the task-based runtime, including task creation, allocation and scheduling.
The point chosen for the small granularity is at around $50\%$ of the peak benchmark performance.
With these two points, we can verify any significant performance difference in peak performance or high
runtime overhead scenarios between the original Nanos6 and our adapted Nanos6+nOS-V version.

\begin{figure}
    \centering
    \includegraphics[width=1\columnwidth]{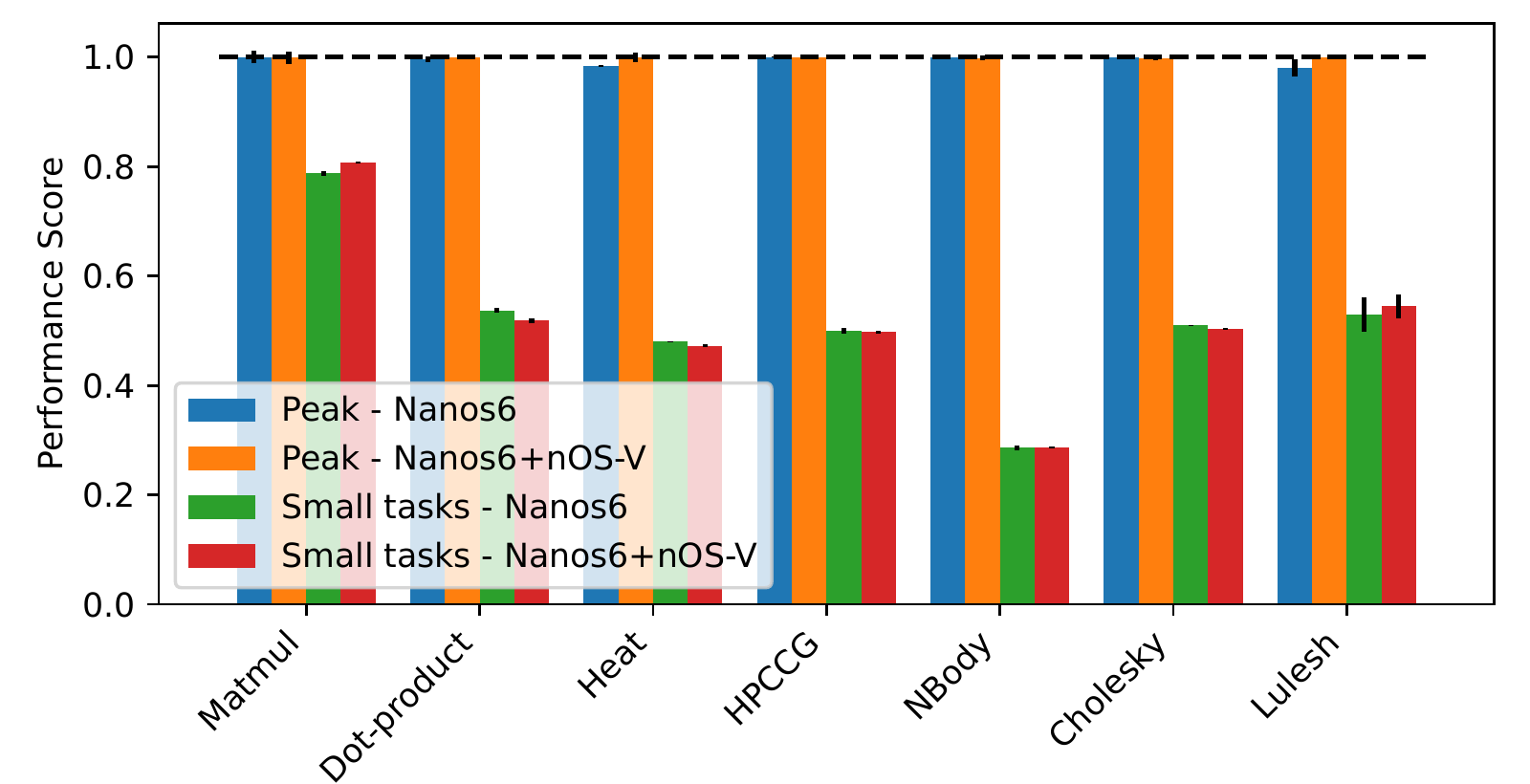}
    \caption{Baseline experiment comparing performance between the original Nanos6 runtime and the Nanos6 runtime adapted with nOS-V support}
    \label{fig:baseline}
\end{figure}

Figure~\ref{fig:baseline} shows the results of the baseline experiment, which are an average of 10 executions using a full node exclusively.
The performance score is calculated as the relative performance versus the best point for each application.
As we can see, there is no relevant speedup or slowdown of the version of Nanos6 with nOS-V when running
a single application aside from minor variances for small granularities.
This experiment confirms that despite using the shared memory regions and adapting our memory allocation to use
shared task descriptors, we have not introduced any relevant performance penalties.

\subsection{Co-execution Performance}

We can now proceed to compare the performance of co-execution versus other node-sharing techniques.
For this experiment, we want to check whether nOS-V can successfully overlap the execution of two and three applications.
The selected applications feature enough parallelism to fill a full node of our test platform.
Some also have sequential initialization phases, reduced parallelism stages, or memory-bound parts where co-location or co-execution could improve overall efficiency.

We used the seven benchmarks from the previous experiment and chose problem sizes to achieve a similar execution time on every benchmark.
Then, we tried six different strategies to execute the chosen applications in all possible pairwise and triple-wise combinations.
The strategies tested were:

\begin{enumerate}
  \item One application after the other (\textbf{exclusive}).
    \item Both applications simultaneously on the whole node, letting the OS manage time-sharing. The Nanos6 runtime blocks idle threads on a futex
      when there are no tasks to execute (\textbf{oversubscription idle}).
	\item Both applications simultaneously on the whole node, letting the OS manage time-sharing, but the Nanos6 runtime performs busy-waiting when there are no tasks
    to execute, to simulate the default configuration of some OpenMP runtimes (\textbf{oversubscription busy}).
  \item Each application on an equal node slice, statically partitioned (\textbf{static co-location}).
  \item Using the DLB library to co-locate applications on the node, using the LeWI policy~\cite{dlb} (\textbf{dynamic co-location}).
  \item Using nOS-V to execute all applications simultaneously (\textbf{co-execution}).
\end{enumerate}

We measured the elapsed time from the start of the application group's execution to when they all finished.
To compare each strategy, we evaluate how close it can get to the minimum achieved makespan of each application combination.

We define $S$ as the set of evaluated strategies, and $A$ as the set of applications used in the evaluation.
With $t_s(x, y)$ as the total makespan for strategy $s \in S$ when running the combination of applications $x \in A$ and $y \in A$, we calculate a \textit{Performance
Score} $p_s(x, y)$ with the following equation:

\[ p_s(x, y) = \frac{\min_{\forall \sigma \in S} t_{\sigma}(x, y)}{t_s(x, y)}\]

We chose this metric instead of speedup to calculate significant averages for each strategy, reflecting how close it can get to be the best performing approach to execute each combination.
Note that the formula is trivially extended for combinations of three applications.

\begin{figure*}
    \centering
    \includegraphics[width=1\textwidth]{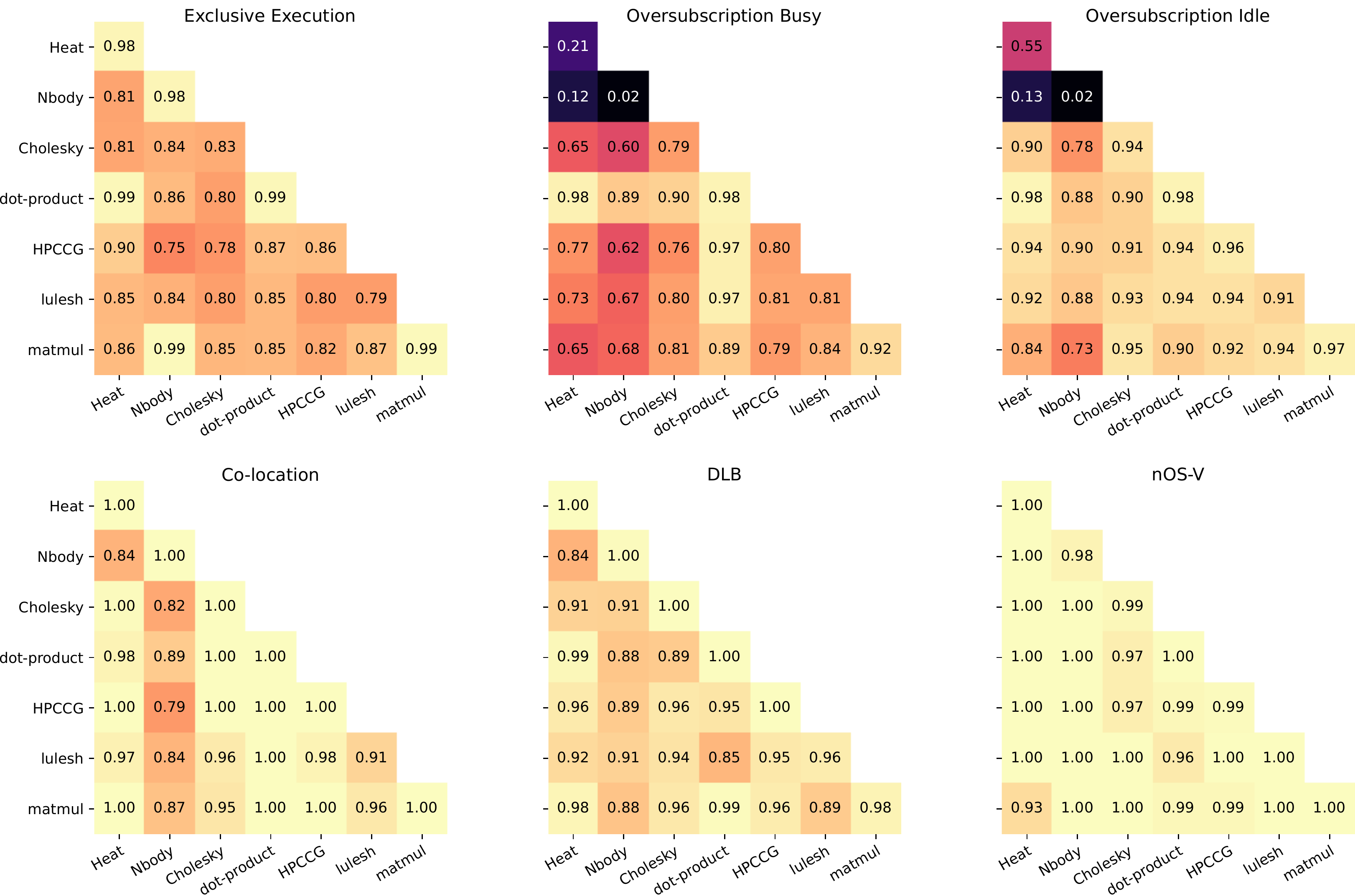}
    \caption{Percentage of peak registered performance achieved for each pairwise benchmark combination on each strategy. Higher is better}
	\label{fig:hmap}
\end{figure*}

Figure \ref{fig:hmap} shows the results of co-scheduling each combination of two applications. The first heatmap displays the performance $p_s(x,y)$ for executing
applications exclusively one after the other, as is the case on most HPC systems. %as has been traditional since the beginning of HPC.
The following heatmaps show the results of applying the proposed strategies.
For both oversubscription-based strategies, the operating system performs time-sharing between the applications.
However, it can clearly cause pathological cases where performance is significantly affected. For example, in the combinations of the N-Body method with the Heat Equation,
or the Cholesky benchmark and the Heat equation.
This showcases the problems we highlighted during the introduction of this paper, where Lock-Holder Preemption incurs large performance penalties on co-scheduled applications.

On the other hand, co-location and DLB can outperform exclusive execution in most co-scheduled workloads for pairwise combinations.
However, there are a couple of cases where the makespan of co-located executions is longer than exclusive execution.
For example, in co-location, the combinations of the Cholesky and N-Body benchmarks and the MatMul and N-Body benchmark cause a slowdown over the exclusive execution version.

Finally, we evaluate the approach proposed in this paper, where we co-execute applications using nOS-V.
In this case, for all application combinations, nOS-V is either the best-performing strategy (with $p_s(x,y) = 1$) or very closely thereafter.
Most importantly, there is no combination where nOS-V performs worse than exclusive execution, which means that co-executing applications
is always beneficial in our experiments, as opposed to all the other evaluated strategies.

% Maybe we could highlight some good results of nOS-V (vs other techniques) of this graphic?

\begin{figure}
    \centering
    \includegraphics[width=1\columnwidth]{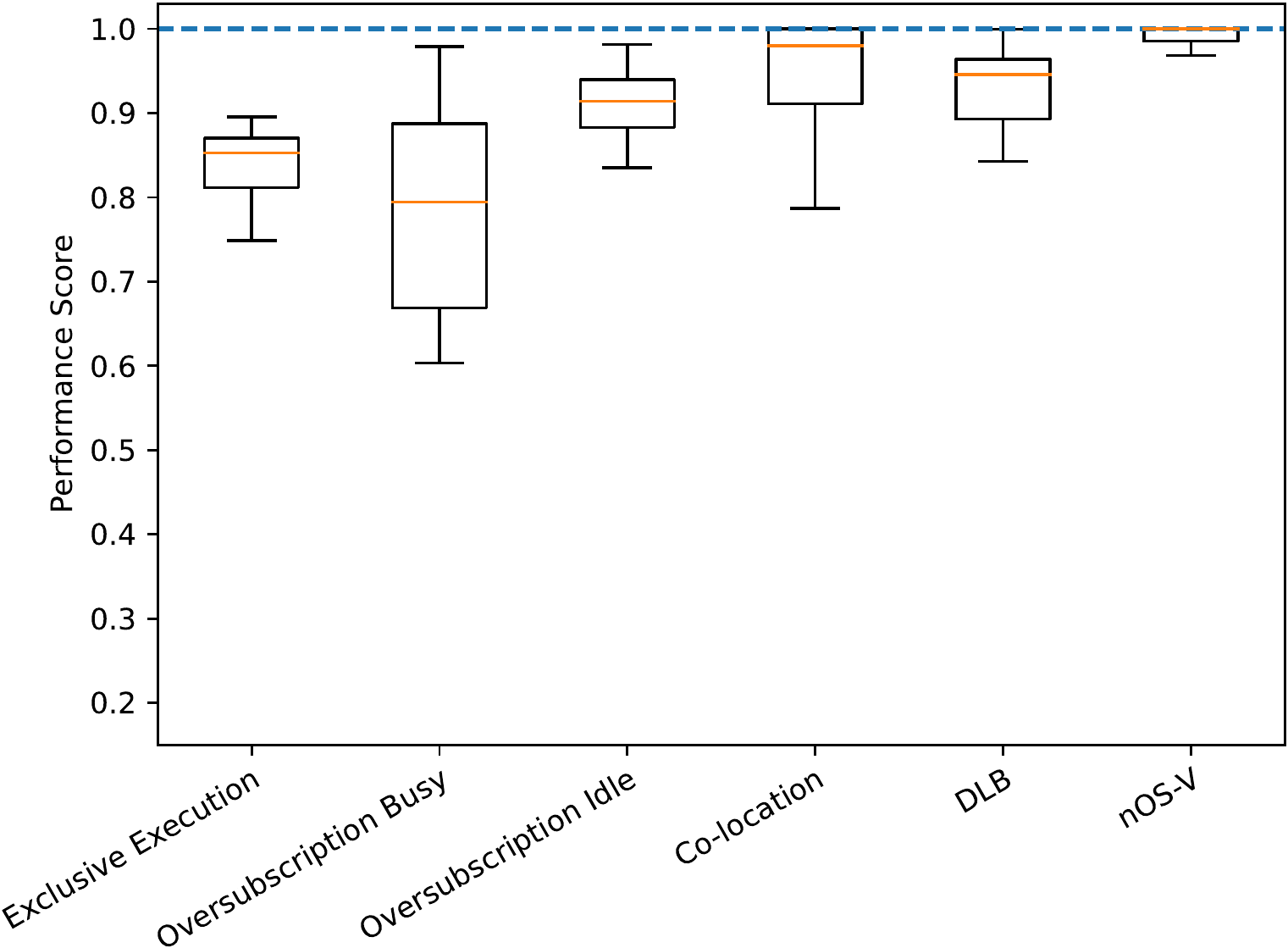}
    \caption{Average performance achieved on pairwise combinations for each strategy, summarized from Figure \ref{fig:hmap}}
	\label{fig:2comb-box}
\end{figure}

Figure \ref{fig:2comb-box} summarizes the scores for all pairwise combinations in every strategy in a box-plot. In this view, it is clear that the best performing
strategy is nOS-V, with a median performance of $1$.
The nOS-V approach also features the smallest Interquartile Range (IQR) and the smallest difference between its \textit{maximum} and \textit{minimum} scores, meaning that its performance score is very near to $1$ in almost all applications' combinations, as shown in Figure \ref{fig:hmap}.
Then, static co-location is the second best performing approach with a median score of $0.979$. However, in this case, there is a significant number of applications' combinations that have lower performance, depicted in the box-plot as higher variability.
The worst technique is the oversubscription busy with a median of $0.78$ but also with the largest variability of scores between combinations.

\begin{figure}
    \centering
    \includegraphics[width=1\columnwidth]{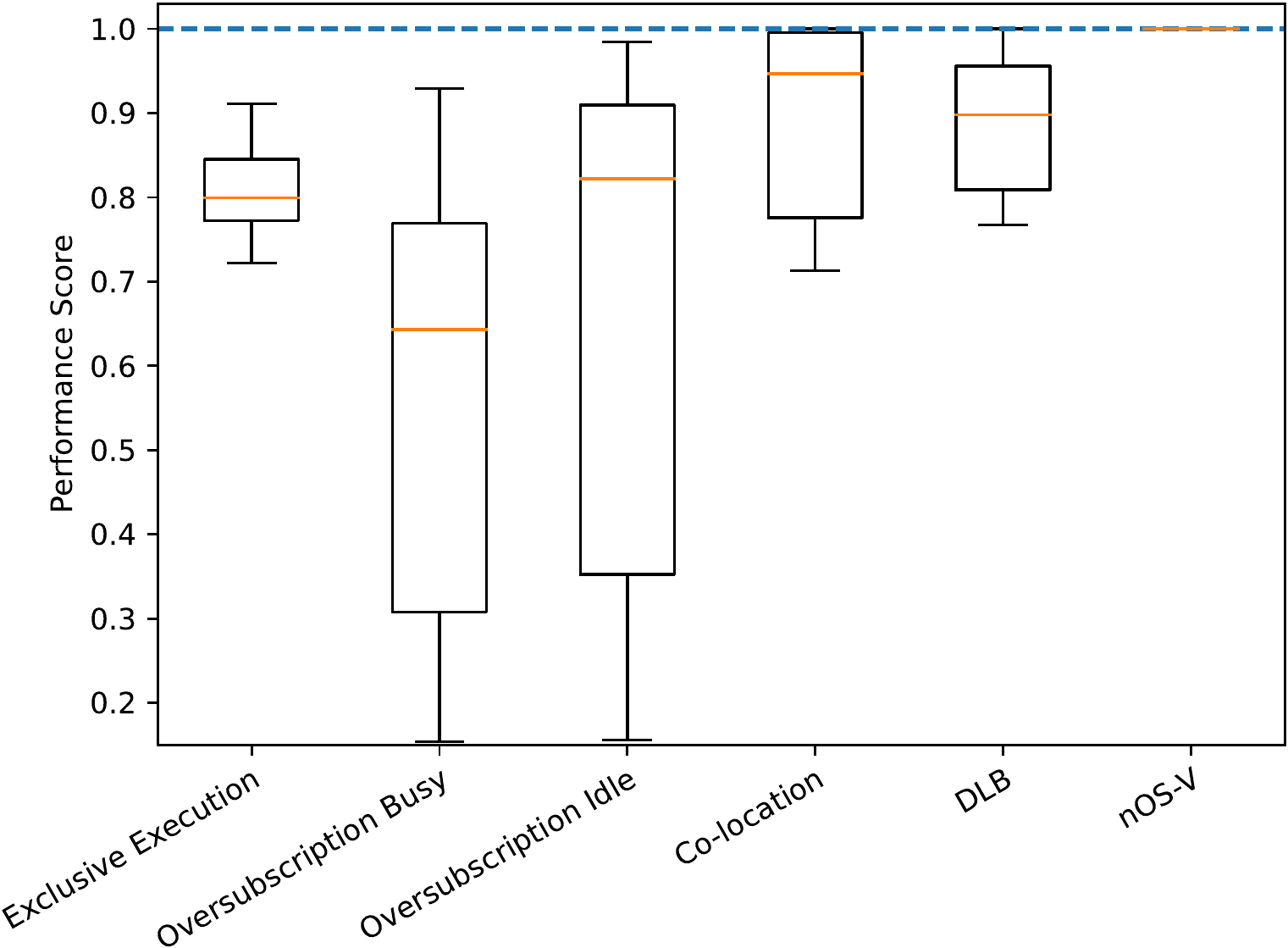}
    \caption{Average performance achieved on three-wise combinations for each strategy. Higher is better}
	\label{fig:3comb-box}
\end{figure}

We then extended the evaluation to co-schedule all three-wise combinations of applications, in this case measuring without repetition.
We executed the resulting 35 possible combinations with all of the evaluated strategies.
Results are reported in Figure \ref{fig:3comb-box}, summarized for all application combinations.
Here, the superiority of the co-execution approach with nOS-V is much clearer.
Median performance is $1$ for all benchmarks and combinations, with very low variance.
For the rest of the approaches, we see that exclusive execution and oversubscription have the lowest medians, as expected. Co-location has a lower median than DLB,
making it the second-best approach, but the variance is significantly higher than DLB.

Note that the median speedup for nOS-V over exclusive execution is $1.17x$ for pairwise combinations, and $1.25x$ for three-wise.
Hence, nOS-V is even more efficient when executing three applications at a time.
This is a key insight, as all other evaluated methods deliver lower speedups versus exclusive execution when we increase the number of co-scheduled applications.

Another relevant question is whether the results achieved by nOS-V are close to a hypothetical, optimal strategy.
To answer this, we can look at two relevant pairwise application combinations: first, the combination which had the lowest
speedup of nOS-V versus the exclusive execution, and then the combination with the maximum attained speedup.

One of the combinations with the lowest speedup was the dot-product's and Heat equation combination.
We can derive the speedup by dividing the performance score of nOSV by the exclusive execution score, in this case a $1x$.
However, if we analyze both applications, we measure CPU utilization of both the dot-product and Heat applications to be $99.5\%$ and
$95.22\%$ respectively when running exclusively, and both benchmarks are memory-bound on our target platform (they have a mean memory bandwidth of $111$ GB/s and $68.95$ GB/s respectively).
This causes a situation where adding more cores to any of the benchmarks without increasing the available memory bandwidth is not productive, and the best possible
result is a uniform splitting of resources, which is why all tested co-scheduling strategies achieve the exact same results.

On the other hand, the highest speedup obtained by nOS-V versus the exclusive execution is a $1.33x$ speedup on the combination of the HPCCG and N-Body applications.
In this case, the CPU utilizations are $73.3\%$ for the HPCCG application and $98.38\%$ for the N-Body, and memory bandwidth usage is $90.21$ GB/s and $0.66$ GB/s respectively.
Our HPCCG is memory-bound, but the N-Body benchmark is compute-bound and uses very little memory bandwidth.
A naive model, knowing both applications have the same execution time $t_o$, would be to assume that they achieve $100\%$ CPU utilization when co-scheduled,
theoretical optimal execution time would be $t_{comb} = (t_o \cdot 73.3 + t_o \cdot 98.38)/100$, the resulting speedup would be $S = (2 \cdot t_o)/t_{comb} = 1.16x$.

However, memory-bound applications can be combined with compute-bound ones because it is generally not needed to use all of the cores of a given platform
to achieve maximum memory bandwidth. In the case of our AMD machine, half of the cores (one per CCX) can fully saturate the chip's bandwidth.
In fact, if we assume that the HPCCG CPU usage can be lowered to $50\%$ without sacrificing performance, the previous formula would result in a $1.35x$ speedup, which
is close to the speedup achieved by nOS-V.

Note that, in any case, obtaining a speedup equal to or higher than $2x$ is infeasible, because it would mean that one of the applications runs faster with less resources,
which is not true for our benchmarks.

These results are very relevant, as we were already working with highly parallel applications that could fill a whole node alone.
Co-execution managed to outperform or closely match all other node-level resource sharing techniques in all application combinations.
Furthermore, as more applications are co-scheduled in the same node, other techniques struggle to provide significant efficiency
gains, as is the case of executing the three benchmarks simultaneously, because it becomes increasingly difficult to find good partitions.

\subsection{Distributed Co-execution and NUMA Architectures}

The previous experiment studied how co-location and co-execution can affect optimized applications that do not run out of parallelism often. % I don't understand the "despite having ..."

This experiment uses hybrid MPI+OmpSs-2 versions of the N-Body simulation and the HPCCG benchmark.
These applications have serial communication phases followed by parallel computation phases that can use all of the available cores, following a Bulk-Synchronous Parallel~\cite{bspc} model.

Additionally, the HPCCG benchmark has a strong NUMA effect. When run in a dual-socket machine, as our Intel Skylake cluster,
the best configuration implies using two MPI ranks per node pinned to each socket. In contrast, the N-Body benchmark is usually run
with one MPI rank per node, as it is strongly compute bound and generally undisturbed by NUMA effects.

As NUMA machines have become increasingly popular in HPC, we designed this experiment to show how nOS-V is uniquely suited to co-schedule
applications in these machines, instead of static or dynamic partitioning.

The problem faced is that when using DLB or nOS-V to co-execute these applications, tasks may migrate from one socket to the other, decreasing overall performance.
If an HPCCG task from Socket 0 is executed in Socket 1, it will perform many remote NUMA accesses.
To prevent this from happening, we can use one of the scheduling policies available in nOS-V thanks to its global scheduler: per-task affinity.

Using the nOS-V API, a runtime such as Nanos6 can configure a preferred or strict affinity for each created task.
Using this API, we can mark locality-sensitive tasks to be executed in the same socket their data resides on, reducing remote NUMA accesses.
To evaluate its effectiveness in tackling NUMA effects, we executed both the N-Body (1 rank per node) and HPCCG (2 ranks per node) applications
using eight nodes of our Intel Skylake cluster.
We measured the makespan of exclusive execution (one after the other, using the full nodes)
versus static co-location, DLB, nOS-V and then nOS-V but pinning the HPCCG tasks to the socket their data resides in.

\begin{figure}
    \centering
    \includegraphics[width=1\columnwidth]{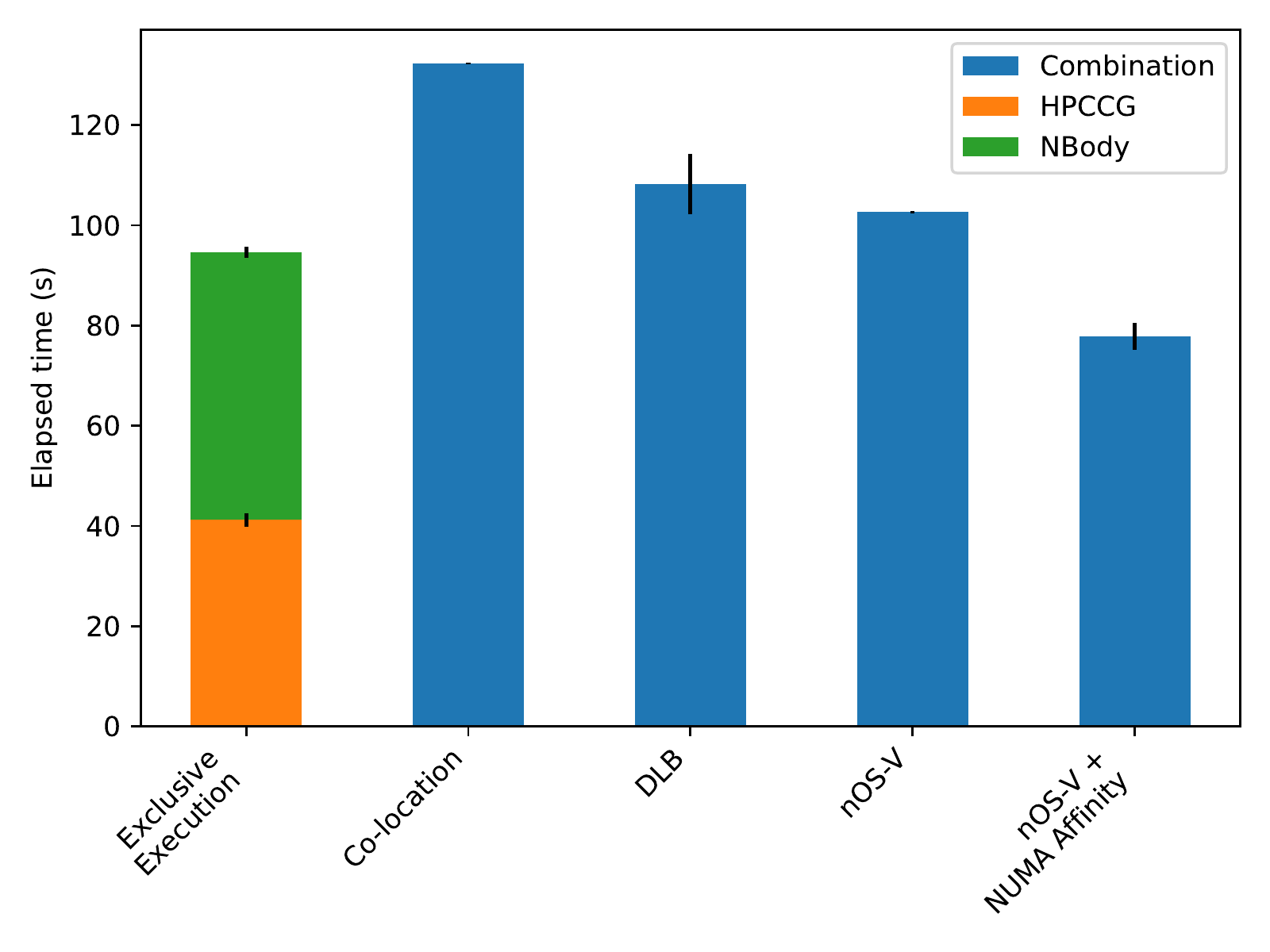}
    \caption{Measuring total makespan of executing the HPCCG and the Nbody distributed applications on the Intel Skylake cluster}
    \label{fig:hpccg-nbody}
\end{figure}

Figure~\ref{fig:hpccg-nbody} shows the result of the experiment.
When running under co-location, statically partitioning the machine in half proved not to be the optimal distribution, resulting in an increased makespan.
This effect can be mitigated using DLB or nOS-V,
but then migrating tasks between sockets causes increased remote accesses, increasing the makespan again.
%Finally, if correctly pin tasks from NUMA-sensitive applications to their home sockets, as shown in the last case, we can obtain a 1.21x speedup over exclusive execution.
Finally, the last bar shows that on NUMA-sensitive applications, if tasks are pinned to the right NUMA node, we can obtain a 1.21x speedup over exclusive execution.

\begin{figure}
    \centering
    \includegraphics[width=1\columnwidth]{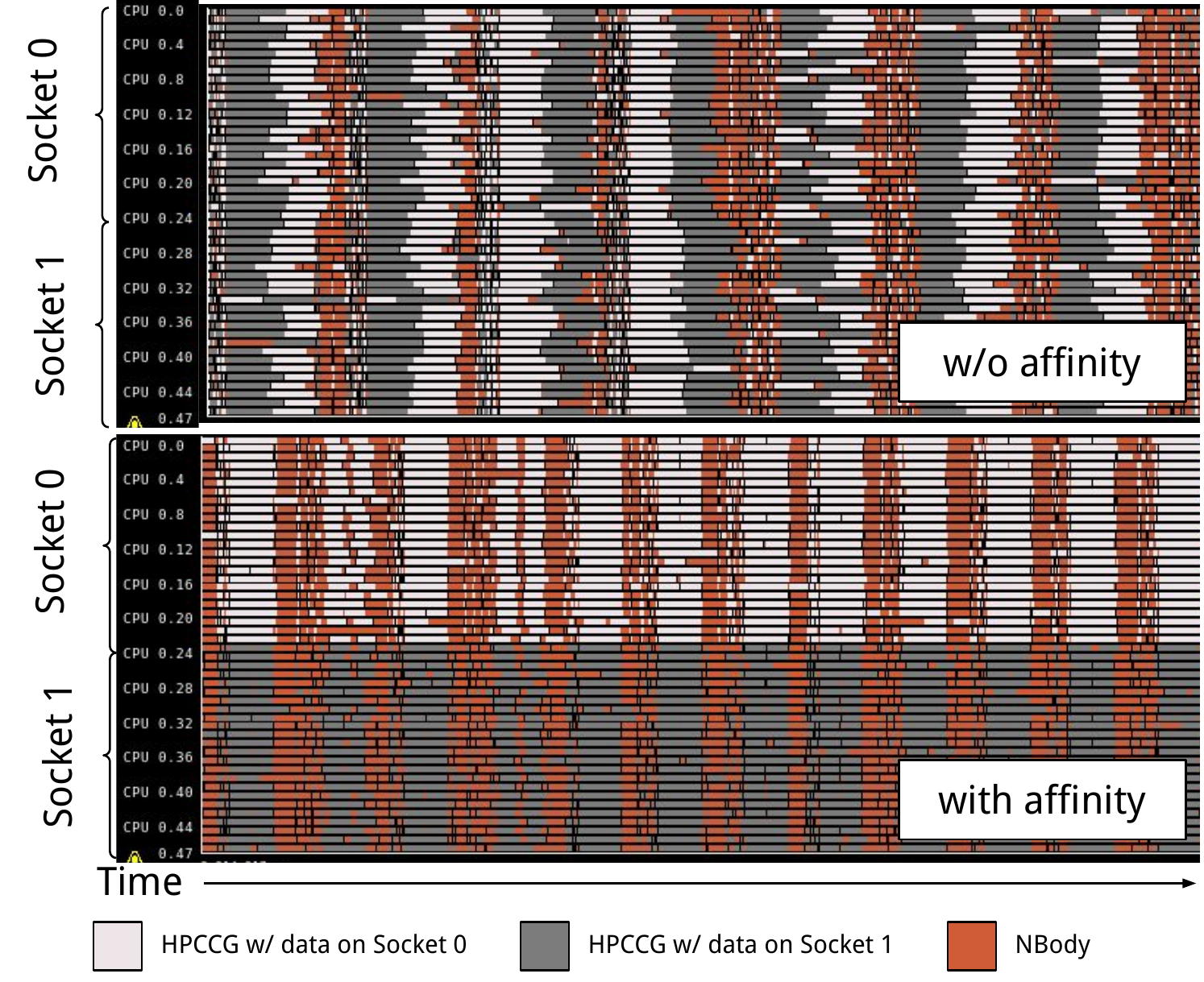}
    \caption{Execution trace of the HPCCG and N-Body benchmarks extracted from a single node of a distributed execution on 8 nodes}
    \label{fig:trace}
\end{figure}

Note that identifying NUMA-sensitive distributed applications and applying affinity to the tasks is often trivial: if an application is
sent to the cluster scheduler with 1-rank per NUMA node, the Nanos6 runtime could automatically pin the tasks to the NUMA node the
rank was assigned, which is the strategy used in this experiment.

We can have a closer look at the affinity policy using the tracing features of nOS-V, which allow us to extract detailed execution traces.
Figure~\ref{fig:trace} shows two execution traces in one node of the HPCCG and N-Body benchmarks used in this experiment.
%In white and gray are both ranks of the HPCCG, and red represents tasks from the N-Body. Each line corresponds to a different CPU, 24 on each socket.
In white and gray are tasks from HPCCG rank 0 and rank 1 respectively, while red represents tasks from the N-Body.
Each line corresponds to a different core, 24 on each socket.
When the applications are executed without any affinity, tasks from both ranks of the HPCCG can execute freely on any socket, resulting in a $70.4\%$ of remote NUMA memory accesses.
However, in the bottom trace, the tasks of the HPCCG rank 0 have their affinity set to socket 0 (cores 0-23), and the tasks from rank 1 to socket 1 (cores 24-47), reducing the tasks' execution time and resulting in a negligible amount of remote NUMA accesses.
\section{Future Work and Challenges}
In this work, we have focused on establishing the fundamentals of co-execution,
and we have showcased the potential benefits and use-cases for this approach.
However, there are some remaining challenges and some possible extensions.

We believe that the applicability of nOS-V is not limited to co-executing
HPC applications, but can be leveraged by any software that can benefit from cooperatively
executing in a shared machine.
As nOS-V exposes a generic user-level threading interface, it can be leveraged by any application that can use user-level threads, even outside of the HPC domain.

There are also many possible extensions for co-execution. For example, to develop automatic mechanisms that can detect the best applications to co-execute to maximize system
throughput and efficiency and integrate this mechanism to cluster job schedulers, similar to what has been proposed for co-location~\cite{slurm-drom, coloc-hpc}.

Finally, we plan to tackle another
interesting research line to extend co-execution to include GPUs and
accelerators, supporting heterogeneous architectures and allowing applications
to co-execute on these devices.

% Related Work
\section{Related Work}

Dynamic Load-Balancing (DLB)~\cite{dlb, drom} is a dynamic co-location library, which acts as an arbitrer between co-located
processes, but lack a global view of the available work on the node.
Colocation-aware libraries such as Arachne~\cite{Arachne} and Hermes~\cite{Hermes} follow this arbiter mechanism.
These libraries feature a separate user process called a core arbiter or core broker. Applications are linked to a library component
that communicates with this arbiter process, and requests cores according to the instantaneous needs of the application.
Note that, exactly like the case for DLB, the arbiter has no global view of the available tasks for each application, only the requested cores,
and therefore cannot take node-wide scheduling decisions.

Callisto~\cite{Callisto} and LIRA~\cite{LIRA} are co-location libraries that improve the dynamic co-location approach by developing an abstraction for runtimes to express
their available parallelism in the form of work-tickets. Additionally, runtimes using Callisto also need to use its
provided synchronization primitives, to prevent preemption while holding locks. This allows a more fine-grained co-location approach, but still
does not provide the same global view a single runtime would have, as information is lost in the work ticket abstraction.

AMCilk~\cite{AMCilk} is a Cilk runtime framework that supports running multiple Cilk applications in the same node
under a single runtime. As applications run under the same runtime, AMCilk does have a global view of the node in the finest granularity.
However, AMCilk requires that all co-scheduled programs are linked into a single binary ahead of time.
This restriction severely limits its applicability in a production environment where co-scheduled programs are usually unknown ahead of time an may change arbitrarily.
Moreover, the AMCilk approach cannot support hybrid MPI applications, which typically require an individual process per MPI rank.
nOS-V is able to provide the same single runtime paradigm without any of the limitations present in AMCilk.

We highlight that the presented evaluation for nOS-V is also unique upon that of previous works, as we evaluate three-wise
combinations and distributed applications with different NUMA policies.

% Conclusions
\section{Conclusions}

This paper presents nOS-V, a lightweight tasking library that supports application
co-execution using node-wide scheduling.
Application co-execution is shown to be a powerful alternative to both static
and dynamic co-location as well as oversubscription to increase throughput in modern
HPC clusters.
The integration with the Nanos6 runtime provides co-execution capabilities to the OmpSs-2 programming
model that are exploited on single-node and distributed benchmarks.
During this evaluation, nOS-V is shown to perform better than the other evaluated node-sharing techniques, obtaining a median $1.17x$ speedup over exclusive execution for pairwise application combinations.
Moreover, when executing three-wise application combinations, the median speedup increases to $1.25x$.

\section{Acknowledgements}
Funding: This project is supported by the Spanish Ministry of Science and Innovation (contract PID2019-107255GB and TIN2015-65316P)
and by the Generalitat de Catalunya (2017-SGR-1414).

\bibliographystyle{ACM-Reference-Format}
\bibliography{ms}

\end{document}